\documentstyle[preprint,aps]{revtex}

\begin{document}
\bibliographystyle{prsty}
\draft

\title{\bf Simultaneous intraportation of many quantum states within the 
quantum computing network}
\author{Mang Feng
\thanks{Email address: feng@mpipks-dresden.mpg.de}} 
\address{$^{1}$Max-Planck Institute for the Physics of Complex Systems,\\
N$\ddot{o}$thnitzer Street 38, D-01187 Dresden, Germany\\
$^{2}$Laboratory of Magnetic Resonance and Atomic and Molecular Physics,\\ 
Wuhan Institute of Physics and Mathematics, Academia Sinica,\\
Wuhan  430071 People's Republic of China}
\date{\today}
\maketitle

\begin{abstract}

A scheme is proposed for simultaneous intraportation of many unknown quantum states within
a quantum computing network. It is shown that our scheme, much different from the teleportation 
in the strict sense, can be very similar to the original teleportation proposal[Phys.Rev.Lett.{\bf 70}
(1993)1895)] and the efficiency of the scheme for quantum state transmission is very high.
The possible applications of our scheme  are also discussed.
\end{abstract}
\vskip 0.1cm
\narrowtext

Quantum computing$^{[1]}$ is an interesting and hot topic in the 
quantum theory, which can treat efficiently some nondeterministic 
polynomial-time problems 
inaccessible for the existing computer, such as factorization of large 
numbers$^{[2]}$, or solve 
some problems more rapidly, e.g., searching a certain item from a large 
disordered system$^{[3]}$, etc. It has been proven that any operation
in the quantum computing can be decomposed into a series of two basic 
operations$^{[4]}$. 
One is controlled-NOT(CN) gate, defined as $|\epsilon_{1}>|\epsilon_{2}>
\rightarrow |\epsilon_{1}>|\epsilon_{1}\oplus\epsilon_{2}>$ with 
$\epsilon_{1,2}=0,1$, and the other is Hadamard gate $\frac {1}{\sqrt{2}}
\pmatrix{1&1 \cr 1&-1}$, which transforms $|0>$ and $|1>$ to 
$\frac {1}{\sqrt{2}}(|0>+|1>)$ and $\frac {1}{\sqrt{2}}(|0>-|1>)$, 
respectively.  As Hadamard gate is a rotation on a single qubit, which is 
easily realized, the physical realization of the CN 
operation is the key to the achievement of an actual quantum computing.
The experimental demonstration of the CN operation on a
trapped ultracold $Be^{+}$ showed that the quantum computing can be
actually carried out after some technical difficulties, such as the
decoherence and ultracold cooling for large quantities of trapped ions, have 
been overcome$^{[5,6]}$.  

The teleportation of an unknown quantum state over arbitrary distance is 
another interesting and hot topic in the quantum theory. It is also a 
striking demonstration of the nonlocal character of quantum states$^{[7]}$, which
starts at a joint measurement on a particle(labelled as particle $a$)
in an unknown state and a particle b being  one-half  of  a  maximally
entangled pair of particles(b and c), and ends at a suitable unitary
rotation of particle c to restore the state of particle $a$ with the help
of two bits of classical message. It has been proven$^{[8,9]}$ that 
the teleportation is a special reversible quantum operation, which hides the 
quantum information within the correlation between the system and the
environment. Recently, the scheme has been extended to the teleportation of 
continued variables$^{[10]}$, and teleportation with GHZ state$^{[11]}$.
Milburn et al$^{[12]}$ demonstrated theoretically the teleportation via a 
two-mode squeezed vacuum state in light of the proposal in Ref.[13],
and the teleportation was also proposed to construct a variety of
quantum gates, associated with other operations$^{[14]}$. As far as we know,
the teleportation has been demonstrated experimentally by using parametric 
down-conversion in interferometric Bell state analyzers and in k-vector 
entanglement$^{[15,16]}$, NMR method$^{[17]}$ and continuous variables of 
electromagnetic field$^{[18]}$.

Since the essence of the teleportation is based on the quantum entanglement 
which is
also the basis of the quantum parallelism in the quantum computing, a 
scheme$^{[19]}$ for achieving teleportation via quantum computing was proposed 
recently. In that scheme, three quantum states are put into a circuit 
consisting of Hadamard operations and CN ones,
in which one(input from channel $a$) is the unknown quantum state needed to be 
teleported and the other two are auxiliary ones (input from channels b and c
respectively) to be entangled. After the entanglement of the states in channels 
$a$ and b(one-half of the entanglement of b and c), one carries out a series 
of CN and Hadamard operations on the state in channel c with the help of the  
quantum information from channels $a$ and b. Finally, the quantum state 
put in the channel $a$ reappears at the output of the channel c. Obviously, 
this scheme is somewhat different from the original scheme proposed by Bennett 
et al${[7]}$. First, there is no need of joint measurement with Bell
basis states. 
Although the authors made some discussions for the measurement on
the channels $a$ and b at the border line of the sender Alice and the
receiver Bob, it is easily found that the final result is irrelative to
these measurements. Secondly, Bob needs not only a unitary rotation to the
quantum state in channel c, but also CN operations to the quantum state with 
the help of the information from channels $a$ and b being control bits.
Particularly the CN operation between the channels $a$-c and b-c implies
that $a$ and b are not arbitrarily far away from c. So this scheme is not
actually referred to the teleportation, but, in some sense, the {\it intraportation} 
which transmits a state from a channel at the sender to another at the receiver 
within a network system.

Nevertheless, this scheme is an interesting extension for the original
teleportation proposal, which enlightens us to make a more efficient 
transmission 
of the unknown states via quantum computing network. In this contribution, we will 
 extend this idea to try to transmit simultaneously many 
unknown quantum states with the circuit similar to Ref.[19]. It can be found that the pure state,
instead of the entangled state required in Ref.[7], can be used as the auxiliary state for the 
state transmission, and  fewer classical messages are needed in our scheme than in Ref.[7]. However, the entanglement
is still essential to our scheme.  A comparison of our scheme with Ref.[7] will be made to 
show the advantages of our scheme. The possible applications of our scheme will be discussed in the secure transmission of 
quantum states as well as the preparation of nonclassical states. 

Suppose first that two quantum states 
$a|1>+b|0>$ and $e|1>+f|0>$ put respectively in channels 1 and 3 will be 
intraported, and the other state $c|1>+d|0>$ in the channel 2 is auxiliary. 
(In this work, we set channels 1, 2 and 3 corresponding to the
the channels from the top to the bottom.) Alice 
need not make any measurement at the position of the 
vertical dotted line denoting the border line of Alice and Bob, but informs 
Bob by the broadcast or telephone with one bits of message
about the values of c and d. If $c=d=\frac {1}{\sqrt{2}}$, Bob will carry out 
a series of corresponding operations, as shown in Fig.1 where and in
the following figures the channels for transmitting the classical message are 
omitted. Then he will obtain $a|1>+b|0>$ and $e|1>+f|0>$ at the outputs of the 
channels 2 and 1 respectively. If $c=0$ and $d=1$, the operation by Bob is shown 
in Fig.2, where 
he obtains $e|1>+f|0>$ and $a|1>+b|0>$ at the outputs of the channels
1 and 3 respectively. Similarly, if $c=1$ and $d=0$, the operation by Bob is 
demonstrated in Fig.3. The later two procedures in fact result in the 
swapping of quantum states, or identity interchange of quantum states$^{[20]}$.

The scheme can be generalized to the intraportation of the unknown quantum states input
respectively from channels 1 and 2, or channels 2 and 3, where the values of e 
and f, or values of $a$ and b should be informed to Bob. The circuit shown in 
Fig.4 is an example in these respects. From Figs. 1-3 we know that  
 there are three kinds of possible outputs for a certain kind of input
under the present operations. For example, for the input case in Fig.1, the 
three kinds of possible outputs are $(e|1>+f|0>)_{1}$$(a|1>+b|0>)_{2}$
$(|1>+|0>)_{3}$, $(|1>+|0>)_{1}(e|1>+f|0>)_{2}$$(a|1>+b|0>)_{3}$, and
$(e|1>+f|0>)_{1}$$(|1>+|0>)_{2}$$(a|1>+b|0>)_{3}$. So If the auxiliary state input 
by Alice is restricted to  three cases, that
is, $c=d=\frac {1}{\sqrt{2}}$, $c=0$ and $d=1$, and $c=1$ and $d=0$, then
there are totally nine different cases for each intraportation, which in fact
constitute the protocol between Bob and Alice for different operations
performed by Bob corresponding to different inputs set by Alice.
Obviously, in the protocol, it is important to make clear from which
channel the auxiliary state is input, since different situation for
input channels of auxiliary states correspond to different operations
Bob should perform.

As we know, three CN operation sequences will result in the
swapping of two quantum states, i.e., $CN_{12}CN_{21}CN_{12}$
$|\Psi>_{1}|\Phi>_{2}=|\Phi>_{1}|\Psi>_{2}$.
So with different group of these operations, three input quantum states
can also reappear at the outputs of different channels via quantum 
computing, as demonstrated in Fig.5 where $\Psi$, $\Phi$, $\theta$
are arbitrary quantum states. But this swapping operation is different from 
our intraportation scheme. The major difference is that, in our scheme of 
intraportation, the quantum information is divided into two parts: one
is related to the entangled state, and the other is the purely classical
information. Moreover, the three input quantum states are fully
entangled before they are sent to the receiver in our scheme, instead
of the swapping operation depending on the interaction between
arbitrary two states of the three input states. However, by means of the
swapping operations, we can simplify above protocol for intraportation, that
is, Bob only needs to consider one kind of output case for each input case.
For example, in Fig.1, Bob chooses the operations to produce 
$(e|1>+f|0>)_{1}$$(a|1>+b|0>)_{2}$$(|1>+|0>)_{3}$, and the other two kinds of outputs
can be obtained via a group of the swapping operations. Therefore, the protocol 
for the present intraportation scheme can be reduced to including only three 
different cases, and the other different output cases are resorted to
the post-intraportation treatment.

The teleportation of the entangled state is also an interesting topic
for the quantum communication. With the original teleportation proposal$^{[7]}$,
it is easily proven that an unknown entangled state can be teleported via a
known entangled state
as well as two bits of classical message. Our scheme simultaneously 
transmitting two pure quantum states, in some sense, also means that it can
intraport entangled states. As shown in Fig.6, two quantum states $c|1>+d|0>$
and $e|1>+f|0>$ are entangled, after the Hadamard gate and CN gate,
to be $\Psi =\frac {1}{\sqrt{2}}(c+d)(e|01>+f|00>)$
$+\frac {1}{\sqrt{2}}(d-c)(e|10>+f|11>)$. With the help of the
auxiliary state $|0>$ and the corresponding operations by Bob, $\Psi$ reappears
at the outputs of channels 1 and 2. In this procedure, the intraportation
only needs a pure state to be auxiliary, and one bits of classical message 
about the auxiliary state from Alice to Bob, which is more efficient 
than that with the original teleportation proposal. Obviously, before
the intraportation, the protocol should be made between Alice and Bob
for Bob's operation corresponding to the different auxiliary state
input by Alice.

The present scheme 
can be readily extended to intraporting simultaneously many unknown quantum 
states and the intraportation of many-particle entangled states, with 
the increase of the channels transmitting the quantum states,  and a little bit
modification of the circuit. For example, for the case of four
channels, we can intraport simultaneously at most three unknown quantum states
via one bit of classical message after the protocol has been made by Alice
and Bob, as shown in Figs.7, 8 and 9 where the auxiliary states were
input from the lowest channel. Comparing these figures with Figs.1-4, 
and by direct deductions for other many-channel situations, we
find that, for the case of N channels, no matter from which channel
the auxiliary state is input, the operations performed by Alice
can be $H_{N-1}CN_{N-1~ N}H_{N-2}CN_{N-2~ N-1}H_{N-3}\cdots CN_{2 3}CN_{1 2}H_{1}$, and
the first five operations by Bob are $CN_{N-1~ N}H_{N}CN_{1 N}H_{1}H_{N}$. 
The other operations by Bob should be performed in terms of the specific
input situation. These characteristics are useful for the production
of the protocol for intraportation. According to above discussion, we
know that, even for the many-channel situation, there are still three 
different cases in the protocol if we use the swapping operations for the
post-intraportation treatment.

As the quantum states are fully entangled before they are reconstructed at 
the output locations, one possible application of our scheme is  the 
secure transmission of the {\it everyday} quantum messages within a quantum
computing network. Suppose an eavesdropper wants to get messages from
the channels  between the sender Alice  and  the distant receiver Bob.  The 
eavesdropper will not succeed even if he eavesdrops the quantum information 
simultaneously from all 
channels, and meanwhile receives the classical message send by Alice,
as long as he do not know the protocol made by Alice and Bob. As the 
eavesdropper is not clear from which channel the auxiliary 
state is input, the probability for his successful eavesdropping
decreases as the increase of the number of the channels, and the action
of the eavesdropper can be detected by Bob from the comparison of the output 
results with the expected results given by the protocol. Although it also 
needs the judgement for whether there exists the eavesdropping, however, different 
from the standard cryptographic scheme$^{[21]}$, our scheme does not involve the 
quantum key distribution process because we have supposed that Alice and Bob are
not much far away from each other. The protocol 
in our scheme can be made by Alice and Bob meeting at a common place.
One may ask: why does not Alice send her quantum message to Bob directly when 
they meet each other? The key point in our scheme is that, once the protocol 
has been made,  Alice can send her quantum messages to Bob efficiently and safely 
at any time when required. They may meet each other once a month or longer to 
change their protocol for keeping secure transmission of the
quantum information in the next several days. Strictly speaking, our scheme is 
less practical for two much distant users than the standard quantum cryptography, 
whereas it might be more convenient and 
practical for the everyday transmission of the quantum information within a limited
quantum computing network in future. Moreover, a by-product of this application is the preparation 
of different quantum states at the output location. By suitably choosing the operations
as well as the coefficients of the input states, some useful quantum states, such as Bell
states$^{[7]}$, Greenberger-Horne-Zeilinger states$^{[22]}$ and so on, can be 
obtained from the output states. For example, in Fig.1, if Bob does not perform 
the last $CN_{23}$, but make a measurement at the channel 3, then he will obtain 
$ea|11>+fb|00>$ or $eb|10>+fa|01>$ from channels 1 and 2 by projecting the output 
state of the channel 3 on $|0>$ or $|1>$.   

A shortcoming of our scheme is the increase of the quantum gates with the 
increase of the number of the channels. In practical application, we may restrict
the number of the channels in terms of the quantity of the quantum message 
to be transmitted. We can also try to integrate some quantum gates, according to 
the regularity of operations referred to above, to reduce the number of the operation. 

In summary, we pointed out that a former teleportation scheme via quantum computing network
is actually an intraportation within a quantum computing network, and along
this idea, we studied how to intraport simultaneously many unknown 
quantum states. As the classical message from Alice to Bob is 
necessary in the scheme, our scheme is more in tune with the original
teleportation proposal than Ref.[19]. Moreover our scheme is also
more efficient than the transmission of unknown quantum states with 
the original teleportation proposal. Different from Ref.[7], however,
the present scheme do not obviously depend on the entanglement characteristic 
of the auxiliary state itself since the pure state can also
act as the auxiliary state, whereas the entanglement is still the heart
of our scheme. So from this viewpoint, the present scheme does not
follow closely the original teleportation proposal. 
On the other hand, for more
coincidence with the original teleportation proposal, we can use two
auxiliary states for the many-channel cases. For example, in Fig.7, we
can set $e=0$ and $f=1$, thus the Hadamard gate and the following CN
operation make the two auxiliary states entangled. Then one part of the
entanglement is correlated with the two unknown quantum states, and
the other one is sent directly to Bob. As a result, Bob can reproduce
in his side the two unknown quantum states by two bits of classical
messages sent from Alice and some corresponding operations performed according 
to the protocol he made with Alice. Obviously, with this change, our scheme 
is well consistent with the proposal in Ref.[7], whereas our
scheme is still more efficient as many unknown quantum states(e.g. two
unknown quantum states in Fig.7)
are transmitted simultaneously by means of two bits of classical message.
From above analysis, we know that, the 
realization of the present scheme at least requires two prerequisites. One is the 
quantum channels between Alice and Bob, whereas with what material to construct these 
quantum channels is still an open question. The other is the 
experimental achievement of the quantum computing with many-qubit. 
As the seven-qubit quantum computing has been carried out via
NMR$^{[23]}$, and many-qubit quantum computing via trapped ions will
be achieved in the near future$^{[24,25]}$, we believe that our scheme will be
helpful for the exploration of efficient transmission of quantum information
in future actual quantum computing network.

The valuable discussion with Prof. Xiwen Zhu and Kelin Gao is highly acknowledged.
The author also thanks Dr.Yurong Jiang for critically checking the deduction in this work.

Note added: After finishing this work, the author was told that a work$^{[26]}$ extending Ref.[19]
along another direction had been carried out in the field of condensed matter physics.

\newpage
\begin{center}{\bf Captions of the figures}\end{center} 

Fig.1 Intraportation of two quantum states, where the auxiliary state
is $|0>+|1>$ and input from the middle channel.

Fig.2 The same as Fig.1, but the auxiliary state is $|0>$. 

Fig.3 The same as Fig.1, but the auxiliary state is $|1>$.

Fig.4 The same as Fig.1, but the auxiliary state $|0>$ is input from the
lowest channel.

Fig.5 Quantum swapping of three quantum states.

Fig.6 Intraportation of an entangled state.

Fig.7 Intraportation of three quantum states, where the auxiliary state
is $|0>+|1>$ and input from the lowest channel.

Fig.8 The same as Fig.7, but the auxiliary state is $|0>$. 

Fig.9 The same as Fig.7, but the auxiliary state is $|1>$.

\newpage
\begin{small}
\begin{picture}(40,100)
\put(0,100){$a|1>+b|0>$ \line(1,0){95}}
\put(0,80){~~~$|1>+|0>$ \line(1,0){15}}
\put(0,60){$e|1>+f|0>$ \line(1,0){75}}
\put(80,90){\line(0,-1){20}}
\put(80,90){\line(1,0){20}}
\put(80,70){\line(1,0){20}}
\put(100,90){\line(0,-1){20}}
\put(85,75){H}
\put(100,80){\line(1,0){100}}
\put(110,80){\circle*{5}}
\put(110,80){\line(0,-1){25}}
\put(110,60){\circle{10}}
\put(140,80){\circle{10}}
\put(140,100){\line(0,-1){25}}
\put(140,100){\circle*{5}}
\put(160,110){\line(0,-1){20}}
\put(160,110){\line(1,0){20}}
\put(160,90){\line(1,0){20}}
\put(180,110){\line(0,-1){20}}
\put(165,95){H}
\put(180,100){\line(1,0){110}}
\put(310,100){\line(1,0){40}}
\put(150,80){\line(1,0){200}}
\put(130,60){\line(1,0){100}}
\put(210,60){\circle{10}}
\put(210,80){\line(0,-1){25}}
\put(210,80){\circle*{5}}
\put(230,70){\line(0,-1){20}}
\put(230,70){\line(1,0){20}}
\put(230,50){\line(1,0){20}}
\put(250,70){\line(0,-1){20}}
\put(235,55){H}

\put(250,60){\line(1,0){40}}
\put(310,60){\line(1,0){40}}
\put(270,100){\circle*{5}}
\put(270,100){\line(0,-1){45}}
\put(270,60){\circle{10}}

\put(290,110){\line(0,-1){20}}
\put(290,110){\line(1,0){20}}
\put(290,90){\line(1,0){20}}
\put(310,110){\line(0,-1){20}}
\put(295,95){H}

\put(290,70){\line(0,-1){20}}
\put(290,70){\line(1,0){20}}
\put(290,50){\line(1,0){20}}
\put(310,70){\line(0,-1){20}}
\put(295,55){H}

\put(330,80){\circle*{5}}
\put(330,80){\line(0,-1){25}}
\put(330,60){\circle{10}}

\put(310,60){\line(1,0){40}}
\put(355,100){$e|1>+f|0>$}
\put(355,80){$a|1>+b|0>$}
\put(355,60){$(|1>+|0>)/\sqrt{2}$}

\put(130,40){Alice}
\put(260,40){Bob}
\put(200,20){Fig.1}
\put(190,110){\circle*{3}}
\put(190,100){\circle*{3}}
\put(190,90){\circle*{3}}
\put(190,80){\circle*{3}}
\put(190,70){\circle*{3}}
\put(190,60){\circle*{3}}
\put(190,50){\circle*{3}}
\end{picture}\\

\begin{picture}(40,80)
\put(0,100){$a|1>+b|0>$ \line(1,0){95}}
\put(0,80){~~~~~$|0>~~~~~~$ \line(1,0){15}}
\put(0,60){$e|1>+f|0>$ \line(1,0){80}}
\put(80,90){\line(0,-1){20}}
\put(80,90){\line(1,0){20}}
\put(80,70){\line(1,0){20}}
\put(100,90){\line(0,-1){20}}
\put(85,75){H}
\put(100,80){\line(1,0){100}}
\put(110,80){\circle*{5}}
\put(110,80){\line(0,-1){25}}
\put(110,60){\circle{10}}
\put(140,80){\circle{10}}
\put(140,100){\line(0,-1){25}}
\put(140,100){\circle*{5}}
\put(160,110){\line(0,-1){20}}
\put(160,110){\line(1,0){20}}
\put(160,90){\line(1,0){20}}
\put(180,110){\line(0,-1){20}}
\put(165,95){H}
\put(180,100){\line(1,0){110}}
\put(310,100){\line(1,0){60}}
\put(150,80){\line(1,0){220}}
\put(130,60){\line(1,0){100}}
\put(210,60){\circle{10}}
\put(210,80){\line(0,-1){25}}
\put(210,80){\circle*{5}}
\put(230,70){\line(0,-1){20}}
\put(230,70){\line(1,0){20}}
\put(230,50){\line(1,0){20}}
\put(250,70){\line(0,-1){20}}
\put(235,55){H}
\put(250,60){\line(1,0){40}}
\put(310,60){\line(1,0){60}}
\put(270,100){\circle*{5}}
\put(270,100){\line(0,-1){45}}
\put(270,60){\circle{10}}
\put(290,70){\line(0,-1){20}}
\put(290,70){\line(1,0){20}}
\put(290,50){\line(1,0){20}}
\put(310,70){\line(0,-1){20}}
\put(295,55){H}
\put(290,110){\line(0,-1){20}}
\put(290,110){\line(1,0){20}}
\put(290,90){\line(1,0){20}}
\put(310,110){\line(0,-1){20}}
\put(295,95){H}
\put(330,100){\circle*{5}}
\put(330,100){\line(0,-1){45}}
\put(330,60){\circle{10}}

\put(310,80){\line(1,0){60}}

\put(375,100){$e|1>+f|0>$}
\put(375,80){$(|1>+|0>)/\sqrt{2}$}
\put(375,60){$a|1>+b|0>$}
\put(130,40){Alice}
\put(260,40){Bob}
\put(200,20){Fig.2}
\put(190,110){\circle*{3}}
\put(190,100){\circle*{3}}
\put(190,90){\circle*{3}}
\put(190,80){\circle*{3}}
\put(190,70){\circle*{3}}
\put(190,60){\circle*{3}}
\put(190,50){\circle*{3}}
\end{picture}\\

\begin{picture}(40,80)
\put(0,100){$a|1>+b|0>$ \line(1,0){95}}
\put(0,80){~~~~~$|1>~~~~~~$ \line(1,0){15}}
\put(0,60){$e|1>+f|0>$ \line(1,0){80}}
\put(80,90){\line(0,-1){20}}
\put(80,90){\line(1,0){20}}
\put(80,70){\line(1,0){20}}
\put(100,90){\line(0,-1){20}}
\put(85,75){H}
\put(100,80){\line(1,0){100}}
\put(110,80){\circle*{5}}
\put(110,80){\line(0,-1){25}}
\put(110,60){\circle{10}}
\put(140,80){\circle{10}}
\put(140,100){\line(0,-1){25}}
\put(140,100){\circle*{5}}
\put(160,110){\line(0,-1){20}}
\put(160,110){\line(1,0){20}}
\put(160,90){\line(1,0){20}}
\put(180,110){\line(0,-1){20}}
\put(165,95){H}

\put(180,100){\line(1,0){110}}
\put(150,80){\line(1,0){80}}
\put(130,60){\line(1,0){100}}
\put(210,60){\circle{10}}
\put(210,80){\line(0,-1){25}}
\put(210,80){\circle*{5}}
\put(230,70){\line(0,-1){20}}
\put(230,70){\line(1,0){20}}
\put(230,50){\line(1,0){20}}
\put(250,70){\line(0,-1){20}}
\put(235,55){H}
\put(250,60){\line(1,0){40}}

\put(230,80){\line(1,0){140}}
\put(250,60){\line(1,0){40}}
\put(270,100){\circle*{5}}
\put(270,100){\line(0,-1){45}}
\put(270,60){\circle{10}}

\put(290,70){\line(0,-1){20}}
\put(290,70){\line(1,0){20}}
\put(290,50){\line(1,0){20}}
\put(310,70){\line(0,-1){20}}
\put(295,55){H}
\put(290,110){\line(0,-1){20}}
\put(290,110){\line(1,0){20}}
\put(290,90){\line(1,0){20}}
\put(310,110){\line(0,-1){20}}
\put(295,95){H}
\put(310,60){\line(1,0){60}}
\put(330,80){\line(1,0){30}}
\put(310,100){\line(1,0){60}}

\put(330,60){\circle*{5}}
\put(330,85){\line(0,-1){25}}
\put(330,80){\circle{10}}
\put(345,100){\circle*{5}}
\put(345,100){\line(0,-1){25}}
\put(345,80){\circle{10}}
\put(355,100){\circle*{5}}
\put(355,100){\line(0,-1){45}}
\put(355,60){\circle{10}}
\put(375,100){$e|1>+f|0>$}
\put(375,80){$(|1>-|0>)/\sqrt{2}$}
\put(375,60){$a|1>+b|0>$}
\put(130,40){Alice}
\put(260,40){Bob}
\put(200,20){Fig.3}
\put(190,110){\circle*{3}}
\put(190,100){\circle*{3}}
\put(190,90){\circle*{3}}
\put(190,80){\circle*{3}}
\put(190,70){\circle*{3}}
\put(190,60){\circle*{3}}
\put(190,50){\circle*{3}}
\end{picture}\\

\begin{picture}(40,80)
\put(0,100){$a|1>+b|0>$ \line(1,0){95}}
\put(0,80){$c|1>+d|0>$ \line(1,0){15}}
\put(0,60){$~~~~~|0>~~~~~~$ \line(1,0){85}}
\put(80,90){\line(0,-1){20}}
\put(80,90){\line(1,0){20}}
\put(80,70){\line(1,0){20}}
\put(100,90){\line(0,-1){20}}
\put(85,75){H}
\put(100,80){\line(1,0){100}}
\put(110,80){\circle*{5}}
\put(110,80){\line(0,-1){25}}
\put(110,60){\circle{10}}
\put(140,80){\circle{10}}
\put(140,100){\line(0,-1){25}}
\put(140,100){\circle*{5}}
\put(160,110){\line(0,-1){20}}
\put(160,110){\line(1,0){20}}
\put(160,90){\line(1,0){20}}
\put(180,110){\line(0,-1){20}}
\put(165,95){H}
\put(180,100){\line(1,0){110}}
\put(310,100){\line(1,0){85}}
\put(150,80){\line(1,0){245}}

\put(130,60){\line(1,0){100}}
\put(210,60){\circle{10}}
\put(210,80){\line(0,-1){25}}
\put(210,80){\circle*{5}}
\put(230,70){\line(0,-1){20}}
\put(230,70){\line(1,0){20}}
\put(230,50){\line(1,0){20}}
\put(250,70){\line(0,-1){20}}
\put(235,55){H}
\put(250,60){\line(1,0){40}}
\put(270,100){\circle*{5}}
\put(270,100){\line(0,-1){45}}
\put(270,60){\circle{10}}
\put(290,70){\line(0,-1){20}}
\put(290,70){\line(1,0){20}}
\put(290,50){\line(1,0){20}}
\put(310,70){\line(0,-1){20}}
\put(295,55){H}
\put(310,60){\line(1,0){60}}
\put(330,80){\circle*{5}}
\put(330,80){\line(0,-1){25}}
\put(330,60){\circle{10}}
\put(335,100){\line(1,0){35}}
\put(350,60){\circle*{5}}
\put(350,60){\line(0,1){25}}

\put(350,80){\circle{10}}
\put(290,110){\line(0,-1){20}}
\put(290,110){\line(1,0){20}}
\put(290,90){\line(1,0){20}}
\put(310,110){\line(0,-1){20}}
\put(295,95){H}
\put(370,70){\line(0,-1){20}}
\put(370,70){\line(1,0){20}}
\put(370,50){\line(1,0){20}}
\put(390,70){\line(0,-1){20}}
\put(375,55){H}
\put(390,100){\line(1,0){5}}
\put(390,60){\line(1,0){5}}
\put(400,100){$|0>$}
\put(400,80){$a|1>+b|0>$}
\put(400,60){$c|1>+d|0>$}
\put(130,40){Alice}
\put(260,40){Bob}
\put(200,20){Fig.4}
\put(190,110){\circle*{3}}
\put(190,100){\circle*{3}}
\put(190,90){\circle*{3}}
\put(190,80){\circle*{3}}
\put(190,70){\circle*{3}}
\put(190,60){\circle*{3}}
\put(190,50){\circle*{3}}
\end{picture}\\

\begin{center}
\begin{picture}(40,80)
\put(0,100){$\Psi$ \line(1,0){125}}
\put(0,80){$\Phi$ \line(1,0){125}}
\put(0,60){$\theta$ \line(1,0){125}}
\put(20,100){\circle*{5}}
\put(20,100){\line(0,-1){25}}
\put(20,80){\circle{10}}
\put(40,80){\circle*{5}}
\put(40,80){\line(0,1){25}}
\put(40,100){\circle{10}}
\put(60,100){\circle*{5}}
\put(60,100){\line(0,-1){25}}
\put(60,80){\circle{10}}
\put(80,80){\circle*{5}}
\put(80,80){\line(0,-1){25}}
\put(80,60){\circle{10}}
\put(100,60){\circle*{5}}
\put(100,60){\line(0,1){25}}
\put(100,80){\circle{10}}
\put(120,80){\circle*{5}}
\put(120,80){\line(0,-1){25}}
\put(120,60){\circle{10}}
\put(140,100){$\Phi$}
\put(140,80){$\theta$}
\put(140,60){$\Psi$}
\put(80,40){Fig.5}
\end{picture}\\
\end{center}

\begin{picture}(100,80)
\put(0,100){$~~~~~|0>~~~~~~$ \line(1,0){95}}
\put(0,80){$c|1>+d|0>$ \line(1,0){15}}
\put(0,60){$e|1>+f|0>$ \line(1,0){80}}
\put(80,90){\line(0,-1){20}}
\put(80,90){\line(1,0){20}}
\put(80,70){\line(1,0){20}}
\put(100,90){\line(0,-1){20}}
\put(85,75){H}
\put(100,80){\line(1,0){100}}
\put(110,80){\circle*{5}}
\put(110,80){\line(0,-1){25}}
\put(110,60){\circle{10}}
\put(130,80){\circle{10}}
\put(130,100){\line(0,-1){25}}
\put(130,100){\circle*{5}}
\put(160,110){\line(0,-1){20}}
\put(160,110){\line(1,0){20}}
\put(160,90){\line(1,0){20}}
\put(180,110){\line(0,-1){20}}
\put(165,95){H}

\put(180,100){\line(1,0){110}}
\put(150,80){\line(1,0){240}}
\put(130,60){\line(1,0){100}}
\put(210,60){\circle{10}}
\put(210,80){\line(0,-1){25}}
\put(210,80){\circle*{5}}
\put(230,70){\line(0,-1){20}}
\put(230,70){\line(1,0){20}}
\put(230,50){\line(1,0){20}}
\put(250,70){\line(0,-1){20}}
\put(235,55){H}
\put(250,60){\line(1,0){40}}
\put(310,60){\line(1,0){50}}
\put(270,100){\circle*{5}}
\put(270,100){\line(0,-1){45}}
\put(270,60){\circle{10}}
\put(310,100){\line(1,0){80}}
\put(290,110){\line(0,-1){20}}
\put(290,110){\line(1,0){20}}
\put(290,90){\line(1,0){20}}
\put(310,110){\line(0,-1){20}}
\put(295,95){H}
\put(290,70){\line(0,-1){20}}
\put(290,70){\line(1,0){20}}
\put(290,50){\line(1,0){20}}
\put(310,70){\line(0,-1){20}}
\put(295,55){H}
\put(310,80){\line(1,0){40}}
\put(330,60){\circle*{5}}
\put(330,60){\line(0,1){25}}
\put(330,80){\circle{10}}
\put(350,100){\circle*{5}}
\put(350,55){\line(0,1){45}}
\put(350,60){\circle{10}}
\put(370,80){\circle*{5}}
\put(370,100){\circle{10}}
\put(370,80){\line(0,1){25}}
\put(350,60){\line(1,0){40}}
\put(395,90){$\}\Psi$}
\put(395,60){$|0>$}
\put(130,40){Alice}
\put(260,40){Bob}
\put(200,20){Fig.6}
\put(190,110){\circle*{3}}
\put(190,100){\circle*{3}}
\put(190,90){\circle*{3}}
\put(190,80){\circle*{3}}
\put(190,70){\circle*{3}}
\put(190,60){\circle*{3}}
\put(190,50){\circle*{3}}
\end{picture}\\

\begin{picture}(40,100)
\put(0,100){$a|1>+b|0>$ \line(1,0){90}}
\put(0,80){$c|1>+d|0>$ \line(1,0){40}}
\put(0,60){$e|1>+f|0>$ \line(1,0){10}}
\put(0,40){~~~$|1>+|0>$ \line(1,0){145}}

\put(75,70){\line(0,-1){20}}
\put(75,70){\line(1,0){20}}
\put(75,50){\line(1,0){20}}
\put(95,70){\line(0,-1){20}}
\put(80,55){H}

\put(95,60){\line(1,0){155}}
\put(175,100){\line(1,0){75}}

\put(105,60){\circle*{5}}
\put(105,60){\line(0,-1){25}}
\put(105,40){\circle{10}}
\put(135,60){\circle{10}}
\put(135,80){\line(0,-1){25}}
\put(135,80){\circle*{5}}
\put(105,90){\line(0,-1){20}}
\put(105,90){\line(1,0){20}}
\put(105,70){\line(1,0){20}}
\put(125,90){\line(0,-1){20}}
\put(110,75){H}
\put(125,80){\line(1,0){185}}
\put(145,100){\circle*{5}}
\put(145,100){\line(0,-1){25}}
\put(145,80){\circle{10}}

\put(155,110){\line(0,-1){20}}
\put(155,110){\line(1,0){20}}
\put(155,90){\line(1,0){20}}
\put(175,110){\line(0,-1){20}}
\put(160,95){H}

\put(200,40){\circle{10}}
\put(200,60){\line(0,-1){25}}
\put(200,60){\circle*{5}}
\put(210,50){\line(0,-1){20}}
\put(210,50){\line(1,0){20}}
\put(210,30){\line(1,0){20}}
\put(230,50){\line(0,-1){20}}
\put(215,35){H}

\put(240,40){\circle{10}}
\put(240,100){\line(0,-1){65}}
\put(240,100){\circle*{5}}

\put(250,49){\line(0,-1){19}}
\put(250,49){\line(1,0){20}}
\put(250,30){\line(1,0){20}}
\put(270,49){\line(0,-1){19}}
\put(255,35){H}
\put(250,70){\line(0,-1){19}}
\put(250,70){\line(1,0){20}}
\put(250,51){\line(1,0){20}}
\put(270,70){\line(0,-1){19}}
\put(255,55){H}
\put(250,110){\line(0,-1){20}}
\put(250,110){\line(1,0){20}}
\put(250,90){\line(1,0){20}}
\put(270,110){\line(0,-1){20}}
\put(255,95){H}
\put(280,40){\circle{10}}
\put(280,100){\line(0,-1){65}}
\put(280,100){\circle*{5}}

\put(295,80){\circle{10}}
\put(295,40){\line(0,1){45}}
\put(295,40){\circle*{5}}

\put(270,100){\line(1,0){15}}
\put(305,100){\line(1,0){55}}
\put(270,40){\line(1,0){90}}
\put(270,60){\line(1,0){90}}
\put(230,40){\line(1,0){20}}

\put(285,110){\line(0,-1){20}}
\put(285,110){\line(1,0){20}}
\put(285,90){\line(1,0){20}}
\put(305,110){\line(0,-1){20}}
\put(290,95){H}

\put(311,90){\line(0,-1){20}}
\put(311,90){\line(1,0){19}}
\put(311,70){\line(1,0){19}}
\put(330,90){\line(0,-1){20}}
\put(315,75){H}
\put(330,80){\line(1,0){30}}
\put(340,60){\circle{10}}
\put(340,55){\line(0,1){25}}
\put(340,80){\circle*{5}}
\put(350,80){\circle{10}}
\put(350,60){\line(0,1){25}}
\put(350,60){\circle*{5}}

\put(365,100){$~~~~|0>$}
\put(365,80){$e|1>+f|0>$}
\put(365,60){$c|1>+d|0>$}
\put(365,40){$a|1>+b|0>$}

\put(130,20){Alice}
\put(270,20){Bob}
\put(200,0){Fig.7}
\put(185,110){\circle*{3}}
\put(185,100){\circle*{3}}
\put(185,90){\circle*{3}}
\put(185,80){\circle*{3}}
\put(185,70){\circle*{3}}
\put(185,60){\circle*{3}}
\put(185,50){\circle*{3}}
\put(185,40){\circle*{3}}
\put(185,30){\circle*{3}}
\end{picture}\\

\begin{picture}(40,100)
\put(0,100){$a|1>+b|0>$ \line(1,0){90}}
\put(0,80){$c|1>+d|0>$ \line(1,0){40}}
\put(0,60){$e|1>+f|0>$ \line(1,0){10}}
\put(0,40){~~~~~~~~~~~$|0>$ \line(1,0){145}}

\put(75,70){\line(0,-1){20}}
\put(75,70){\line(1,0){20}}
\put(75,50){\line(1,0){20}}
\put(95,70){\line(0,-1){20}}
\put(80,55){H}

\put(95,60){\line(1,0){155}}
\put(175,100){\line(1,0){75}}

\put(105,60){\circle*{5}}
\put(105,60){\line(0,-1){25}}
\put(105,40){\circle{10}}
\put(135,60){\circle{10}}
\put(135,80){\line(0,-1){25}}
\put(135,80){\circle*{5}}
\put(105,90){\line(0,-1){20}}
\put(105,90){\line(1,0){20}}
\put(105,70){\line(1,0){20}}
\put(125,90){\line(0,-1){20}}
\put(110,75){H}
\put(125,80){\line(1,0){185}}
\put(145,100){\circle*{5}}
\put(145,100){\line(0,-1){25}}
\put(145,80){\circle{10}}

\put(155,110){\line(0,-1){20}}
\put(155,110){\line(1,0){20}}
\put(155,90){\line(1,0){20}}
\put(175,110){\line(0,-1){20}}
\put(160,95){H}

\put(200,40){\circle{10}}
\put(200,60){\line(0,-1){25}}
\put(200,60){\circle*{5}}
\put(210,50){\line(0,-1){20}}
\put(210,50){\line(1,0){20}}
\put(210,30){\line(1,0){20}}
\put(230,50){\line(0,-1){20}}
\put(215,35){H}

\put(240,40){\circle{10}}
\put(240,100){\line(0,-1){65}}
\put(240,100){\circle*{5}}

\put(250,49){\line(0,-1){19}}
\put(250,49){\line(1,0){20}}
\put(250,30){\line(1,0){20}}
\put(270,49){\line(0,-1){19}}
\put(255,35){H}
\put(250,70){\line(0,-1){19}}
\put(250,70){\line(1,0){20}}
\put(250,51){\line(1,0){20}}
\put(270,70){\line(0,-1){19}}
\put(255,55){H}
\put(250,110){\line(0,-1){20}}
\put(250,110){\line(1,0){20}}
\put(250,90){\line(1,0){20}}
\put(270,110){\line(0,-1){20}}
\put(255,95){H}

\put(280,100){\circle{10}}
\put(280,80){\line(0,1){25}}
\put(280,80){\circle*{5}}

\put(295,80){\circle{10}}
\put(295,40){\line(0,1){45}}
\put(295,40){\circle*{5}}

\put(270,100){\line(1,0){90}}
\put(270,60){\line(1,0){90}}
\put(270,40){\line(1,0){35}}
\put(230,40){\line(1,0){20}}
\put(325,40){\line(1,0){35}}

\put(305,50){\line(0,-1){20}}
\put(305,50){\line(1,0){20}}
\put(305,30){\line(1,0){20}}
\put(325,50){\line(0,-1){20}}
\put(310,35){H}

\put(310,80){\line(1,0){50}}

\put(330,60){\circle{10}}
\put(330,40){\line(0,1){25}}
\put(330,40){\circle*{5}}

\put(345,40){\circle{10}}
\put(345,35){\line(0,1){25}}
\put(345,60){\circle*{5}}

\put(365,100){$~~~~|0>$}
\put(365,80){$a|1>+b|0>$}
\put(365,60){$c|1>+d|0>$}
\put(365,40){$e|1>+f|0>$}

\put(130,20){Alice}
\put(270,20){Bob}
\put(200,0){Fig.8}
\put(185,110){\circle*{3}}
\put(185,100){\circle*{3}}
\put(185,90){\circle*{3}}
\put(185,80){\circle*{3}}
\put(185,70){\circle*{3}}
\put(185,60){\circle*{3}}
\put(185,50){\circle*{3}}
\put(185,40){\circle*{3}}
\put(185,30){\circle*{3}}
\end{picture}\\

\begin{picture}(40,100)
\put(0,100){$a|1>+b|0>$ \line(1,0){90}}
\put(0,80){$c|1>+d|0>$ \line(1,0){40}}
\put(0,60){$e|1>+f|0>$ \line(1,0){10}}
\put(0,40){~~~~~~~~~~~$|1>$ \line(1,0){145}}
\put(75,70){\line(0,-1){20}}
\put(75,70){\line(1,0){20}}
\put(75,50){\line(1,0){20}}
\put(95,70){\line(0,-1){20}}
\put(80,55){H}
\put(95,60){\line(1,0){155}}
\put(175,100){\line(1,0){75}}
\put(105,60){\circle*{5}}
\put(105,60){\line(0,-1){25}}
\put(105,40){\circle{10}}
\put(135,60){\circle{10}}
\put(135,80){\line(0,-1){25}}
\put(135,80){\circle*{5}}
\put(105,90){\line(0,-1){20}}
\put(105,90){\line(1,0){20}}
\put(105,70){\line(1,0){20}}
\put(125,90){\line(0,-1){20}}
\put(110,75){H}
\put(125,80){\line(1,0){185}}
\put(145,100){\circle*{5}}
\put(145,100){\line(0,-1){25}}
\put(145,80){\circle{10}}
\put(155,110){\line(0,-1){20}}
\put(155,110){\line(1,0){20}}
\put(155,90){\line(1,0){20}}
\put(175,110){\line(0,-1){20}}
\put(160,95){H}
\put(200,40){\circle{10}}
\put(200,60){\line(0,-1){25}}
\put(200,60){\circle*{5}}
\put(210,50){\line(0,-1){20}}
\put(210,50){\line(1,0){20}}
\put(210,30){\line(1,0){20}}
\put(230,50){\line(0,-1){20}}
\put(215,35){H}
\put(240,40){\circle{10}}
\put(240,100){\line(0,-1){65}}
\put(240,100){\circle*{5}}
\put(250,49){\line(0,-1){19}}
\put(250,49){\line(1,0){20}}
\put(250,30){\line(1,0){20}}
\put(270,49){\line(0,-1){19}}
\put(255,35){H}
\put(250,70){\line(0,-1){19}}
\put(250,70){\line(1,0){20}}
\put(250,51){\line(1,0){20}}
\put(270,70){\line(0,-1){19}}
\put(255,55){H}
\put(250,110){\line(0,-1){20}}
\put(250,110){\line(1,0){20}}
\put(250,90){\line(1,0){20}}
\put(270,110){\line(0,-1){20}}
\put(255,95){H}
\put(280,40){\circle{10}}
\put(280,100){\line(0,-1){65}}
\put(280,100){\circle*{5}}
\put(290,80){\circle{10}}
\put(290,40){\line(0,1){45}}
\put(290,40){\circle*{5}}
\put(300,100){\circle{10}}
\put(300,80){\line(0,1){25}}
\put(300,80){\circle*{5}}
\put(270,100){\line(1,0){35}}
\put(305,100){\line(1,0){65}}
\put(270,40){\line(1,0){100}}
\put(270,60){\line(1,0){100}}
\put(230,40){\line(1,0){20}}
\put(310,90){\line(0,-1){20}}
\put(310,90){\line(1,0){20}}
\put(310,70){\line(1,0){20}}
\put(330,90){\line(0,-1){20}}
\put(315,75){H}
\put(330,80){\line(1,0){40}}
\put(340,100){\circle{10}}
\put(340,40){\line(0,1){65}}
\put(340,40){\circle*{5}}
\put(350,60){\circle{10}}
\put(350,55){\line(0,1){25}}
\put(350,80){\circle*{5}}
\put(360,80){\circle{10}}
\put(360,60){\line(0,1){25}}
\put(360,60){\circle*{5}}
\put(375,100){$~~~~|1>$}
\put(375,80){$e|1>+f|0>$}
\put(375,60){$c|1>+d|0>$}
\put(375,40){$a|1>+b|0>$}
\put(130,20){Alice}
\put(270,20){Bob}
\put(200,0){Fig.9}
\put(185,110){\circle*{3}}
\put(185,100){\circle*{3}}
\put(185,90){\circle*{3}}
\put(185,80){\circle*{3}}
\put(185,70){\circle*{3}}
\put(185,60){\circle*{3}}
\put(185,50){\circle*{3}}
\put(185,40){\circle*{3}}
\put(185,30){\circle*{3}}
\end{picture}\\

\end{small}
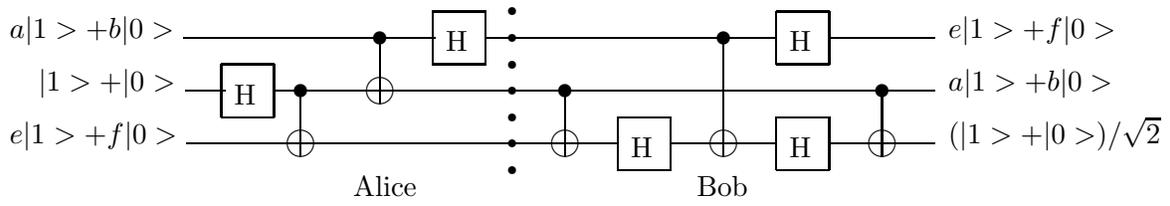
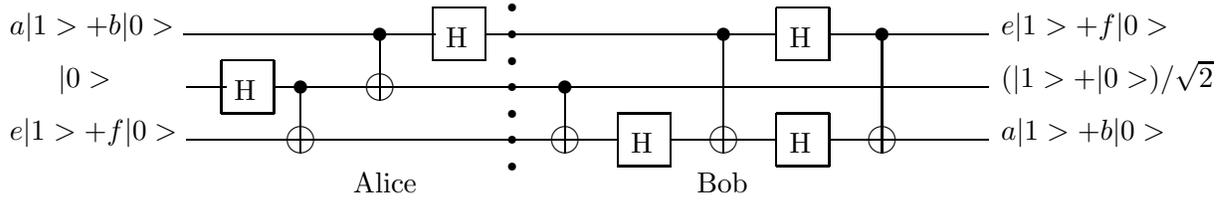
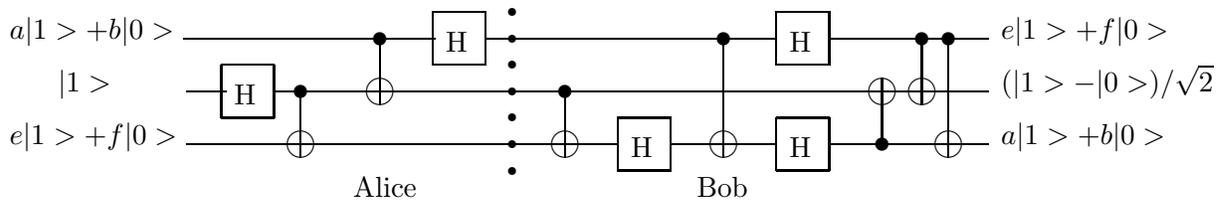
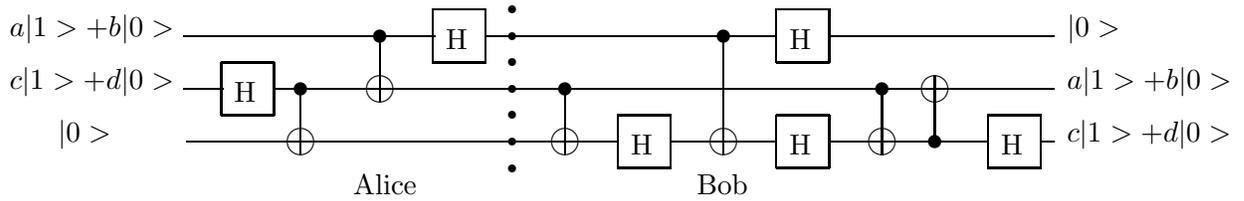
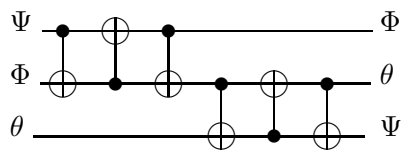
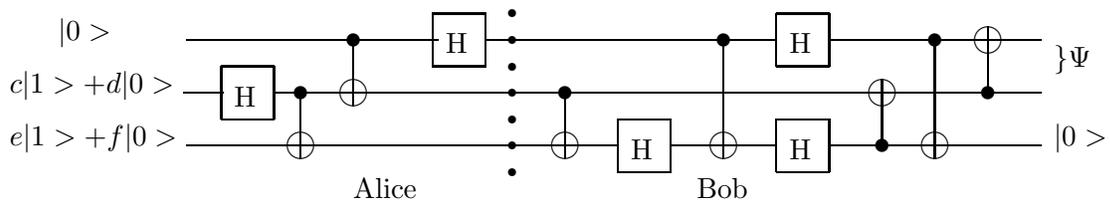
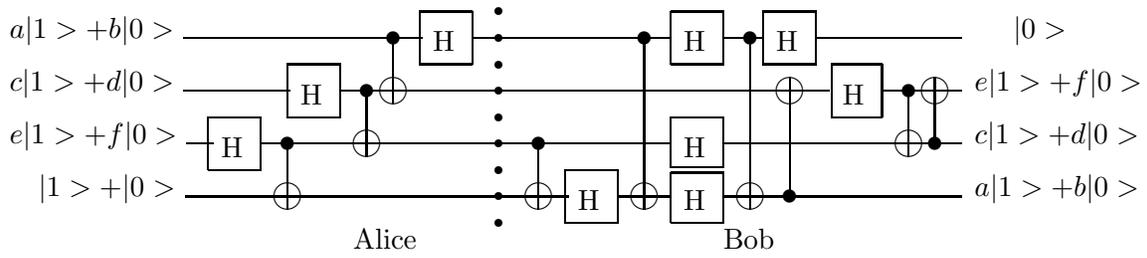
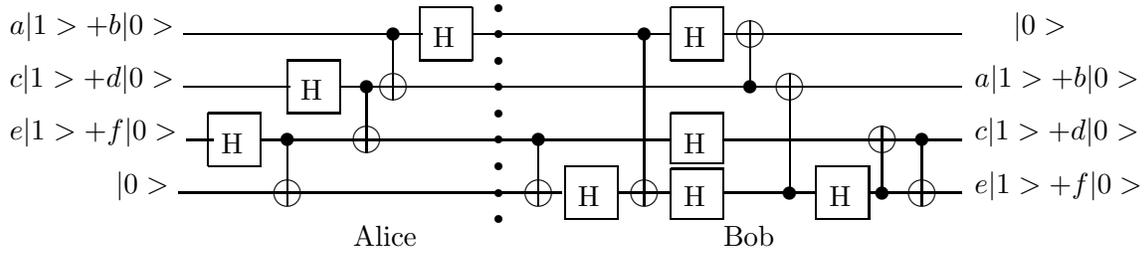
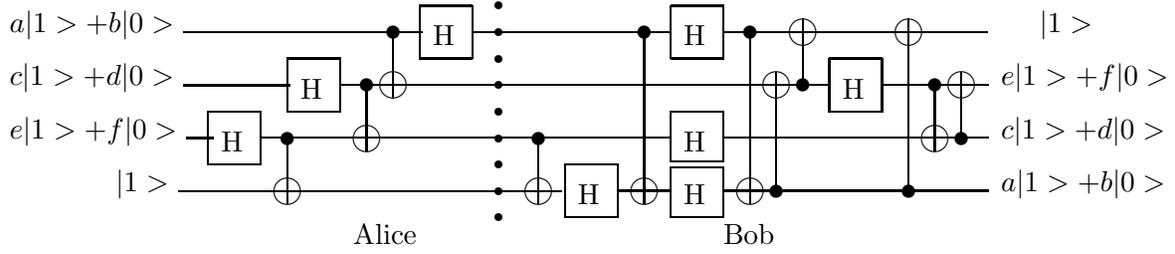
\end{document}